\documentclass[prb,twocolumn,amsmath,amssymb,10pt,aps,longbibliography,superscriptaddress,citeautoscript,bibnotes,footinbib]{revtex4-1}
\usepackage{feynmp}
\usepackage{amsmath}
\usepackage{graphicx}
\usepackage{multirow}
\usepackage{latexsym}
\usepackage{textcomp}
\usepackage{verbatim}
\usepackage{color}
\usepackage{bm}
\usepackage{subfigure}
\usepackage{everysel}
\usepackage{keyval}
\usepackage{ragged2e}
\usepackage{dsfont}
\usepackage{amssymb}
\usepackage{enumitem}
\usepackage{pstricks}
\usepackage{mathtools}
\usepackage{braket}
\usepackage{slashed} 
\usepackage[colorlinks=true]{hyperref}

\usepackage{graphicx}
\usepackage{xcolor}
\usepackage{amsmath}
\usepackage{amsfonts}
\usepackage{bbm}

\usepackage{ulem}

\newcommand{\osum}{{%
    \setbox0\hbox{\circ}%
    \rlap{\hbox to \wd0{\hss\sum\hss}}\box0
}}

\newcommand{\tr}{\text{tr}}

\newcommand{\GYC}[1]{{\color{blue} #1}}
\newcommand{\BK}[1]{{\color{red}  #1}}

\begin{document}

\title{Many-Body Invariants for Chern and Chiral Hinge Insulators}

\author{Byungmin Kang}
\thanks{Electronic Address: bkang119@kias.re.kr}
\affiliation{School of Physics, Korea Institute for Advanced Study, Seoul 02455, Korea}

\author{Wonjun Lee}
\thanks{Electronic Address: wonjun1998@postech.ac.kr}
\affiliation{Department of Physics, Pohang University of Science and Technology (POSTECH), Pohang 37673, Republic of Korea}

\author{Gil Young Cho}
\thanks{Electronic Address: gilyoungcho@postech.ac.kr}
\affiliation{Department of Physics, Pohang University of Science and Technology (POSTECH), Pohang 37673, Republic of Korea}

\date{\today}
 
\begin{abstract} 
We construct new many-body invariants for 2d Chern and 3d chiral hinge insulators, which are characterized by quantized pumping of dipole and quadrupole moments. The invariants that we devise are written entirely in terms of many-body ground state wavefunctions on a torus geometry with a set of unitary operators. We provide a number of supporting evidences for our invariants via topological field theory interpretation, adiabatic pumping argument, and direct mapping to free-fermion band indices. We finally confirm our invariants by numerical computations including infinite density matrix renormalization group on a quasi-one-dimensional system. The many-body invariants therefore explicitly encircle several different pillars of theoretical descriptions of topological phases.
\end{abstract}

\maketitle
\textbf{1. Introduction}: The discovery of topological band insulators initiated a serious revisit to the band theory of insulators, which has fruited in the last fifteen years~\cite{RevModPhys.82.3045, RevModPhys.83.1057}. So far, the classification of the band topology has made a remarkable success and found numerous different kinds of topological insulators~\cite{kitaev2009periodic, RevModPhys.88.035005, bradlyn2017topological}. Insightful band indices and their underlying structures~\cite{RevModPhys.82.3045, RevModPhys.83.1057, benalcazar2017quantized} have been revealed as well. One particular progress, which attracted a huge attention recently, is the so-called higher-order topological insulators~\cite{benalcazar2017quantized, PhysRevB.96.245115, schindler2018higher}, whose topology manifests as the symmetry-protected corner or hinge states instead of more familiar surface states. However, the understanding of these insulators has one important missing conceptual ingredient compared to their ancestors, e.g., insulators with bulk polarization~\cite{PhysRevLett.80.1800} and Chern insulators~\cite{PhysRevLett.61.2015}. For those insulators, we know how to define and detect the topology even without referring to the band structure. For instance, the Resta's formula~\cite{PhysRevLett.80.1800} allows us to measure the polarization in a non-perturbative fashion for any insulators with a unique ground state.

In this Letter, we propose many-body invariants for prototypical chiral topological phases, i.e., the 2d Chern insulator~\cite{PhysRevLett.61.2015} and the 3d chiral hinge insulator~\cite{benalcazar2017quantized, PhysRevB.96.245115, schindler2018higher}. The two insulators are all characterized by quantized chiral pumping of certain quantum numbers under the adiabatic change of the background $U(1)$ gauge field. We show how the topology of such chiral pumping can be detected within the many-body wavefunctions by generalizing Resta's pioneering work~\cite{PhysRevLett.80.1800}. We combine a few different theoretical approaches to support the validity of our invariants: many-body adiabatic pumping argument, topological field theory interpretation, and reduction to free-fermion band indices. We finally confirm our invariants in several numerical calculations. 

%For the Chern insulator, we present a new compact formula that evaluates the Chern number of a given wavefunction, only with the two ground state wavefunctions. This formula can be viewed as the completion of the previous attempts. 

The many-body invariants~\cite{PhysRevLett.80.1800, PhysRevB.100.245134, PhysRevB.100.245135, PhysRevLett.123.196402} can be applied to correlated states of matter~\cite{PhysRevB.88.155121} and to systems which do not have translation symmetries. In practice, such invariants will be useful in the studies on the disordered topological insulators with/without interactions. Even for the clean, free systems, the invariants can provide a complementary diagnosis to the momentum-space, free-electron band indices~\cite{he2019quadrupole, ghosh2019engineering, zeng2019higher, PhysRevLett.124.036803}. Not only practically but also fundamentally, the construction of many-body invariants is important on its own, as they can potentially provide a classification of the quantum phases of matter in general setups.

%, which is perhaps one of the most important goals of modern condensed matter physics. 

%\GYC{By passing, we note that recently there have been several attempts in defining the many-body invariants for higher-order topological insulators, which studied 2d higher-order topology~\cite{PhysRevLett.123.196402, PhysRevResearch.2.012009, You2020higher} instead of the chiral states that we consider here.}

%Generic Introduction: studies on many-body invariants, advantage of real-space expressions, etc. 

%To achieve the goal, 
To this end, we will consider the functions in terms of many-body ground state wavefunctions:  
\begin{align}
\mathcal{Z}[\Phi_a] = \frac{\langle \text{GS}[\Phi_a]| \hat{U}_{\text{top}} |\text{GS}[\Phi_a] \rangle}{\langle \text{GS}[0]| \hat{U}_{\text{top}}|\text{GS}[0] \rangle}, 
\label{Inv}
\end{align}
where we use \textit{two and only two} many-body ground states $|\text{GS}[\Phi_a] \rangle$ on a torus, whose boundary condition along the $a$-direction of the torus is twisted by a flux $\Phi_a$, whereas other directions are chosen to be periodic without any twisting. Here we will choose $\hat{U}_{\text{top}}$ depending on the topology that we want to detect. We will then show that the U(1) phase factor of $\mathcal{Z}[\Phi_a]$ gives the corresponding invariants. 

For the systems that we are interested in, it is sufficient to consider only two unitaries: Resta's polarization operator along the $j$-direction~\cite{PhysRevLett.80.1800},
\begin{align}
\hat{U}_{1,j} = \exp \Big( \frac{2\pi i}{L_j} \sum_{\bm{r}} x_j \hat{n}(\bm{r}) \Big), 
\end{align} 
and the quadrupole operator in the $xy$-plane from our previous work~\cite{PhysRevB.100.245134, PhysRevB.100.245135} 
\begin{align}
\hat{U}_{2} = \exp \Big( \frac{2\pi i}{L_x L_y} \sum_{\bm{r}} xy \hat{n}(\bm{r}) \Big) .
\end{align} 
We assume that these two unitary operators can measure the dipole and the quadrupole moments of many-body insulators under appropriate conditions. We review some of their properties in SI~\cite{suppl}.

\textbf{2. Chern Insulator}: We first construct the formula for the Chern number from the many-body wavefunction $|\text{GS}[\Phi_x] \rangle$ defined on a torus. This will serve as a pedagogical example for our development Eq.~\eqref{Inv}. Here the flux $\Phi_x$ along the $x$-direction is chosen to be small. It can be encoded as the uniform gauge field $A_x = \frac{\Phi_x}{L_x}$. Then we can obtain the \textit{many-body} Chern number $C$ by 
\begin{align}
\mathcal{Z}_c = \frac{\langle \text{GS}[\Phi_x]| \hat{U}_{1,y} |\text{GS}[\Phi_x] \rangle}{\langle \text{GS}[0]| \hat{U}_{1,y} |\text{GS}[0] \rangle} = |\mathcal{Z}|\exp\Big(iC\Phi_x \Big).
\label{Chern}
\end{align}
Since our expression only involves two ground states, it would be useful, e.g., when numerically computing the Chern number of the ground state of the interacting Hamiltonians.

%In an interacting and/or translation non-symmetric system, one can construct an effective Brillouin zone (EBZ) using the the twisted boundary conditions along $x$- and $y$-direction following the idea of Ref.~\onlinecite{PhysRevB.31.3372}. If we assume that the ground state is always gapped and unique over the whole EBZ, the quantized Chern number can be computed by integrating the Berry curvature associated with the ground state over the EBZ, which is the Niu-Thouless-Wu formula~\cite{PhysRevB.31.3372}. This is further simplified to a one-plaqutte formula introduced in Ref.~\onlinecite{PhysRevLett.122.146601}. For translationally invariant interacting fermionic systems with a gap for the ground state near the origin of the EBZ, there exists a remarkable theorem by Hastings and Michalakis~\cite{hastings2015quantization}, who showed that under such setting, the one-plaqutte formula indeed gives the quantized Hall conductivity, a quantity which equals the Chern number in a proper unit. This requires at least 3 ground states to build a single plaqutte. 

%Since constructing a single plaqutte requires at least 3 ground states, a naive application of the theorem by Hastings and Michalakis cannot reduce the number of required ground states to 2.

We first prove that $C$ in Eq.~\eqref{Chern} reduces to the band Chern number when we apply it to a free-fermion band insulator. Here, we only sketch the key steps and defer a more complete derivation to SI~\cite{suppl}. In a free fermion system, we first construct the single-particle eigenstates $|\bm{k}\rangle \otimes |u (\bm{k}) \rangle$, where $|\bm{k}\rangle$ is the plane wave labeled by the (Bloch) momentum and $|u(\bm{k})\rangle$ is the corresponding Bloch state. The many-body ground state is given by the Slater determinant of the occupied single-particle states. Note that the flux $\Phi_x$ merely modifies the allowed momentum $\bm{k}$. So Eq.~\eqref{Chern} for the case of a single filled band insulator becomes 
\begin{align}
\mathcal{Z}_c = \frac{\prod_{k_x, k_y} \langle u (k_x + \frac{\Phi_x}{L_x}, k_y)|u(k_x + \frac{\Phi_x}{L_x}, k_y + \frac{2\pi}{L_y})\rangle}{\prod_{ k_x,k_y}\langle u (k_x, k_y)|u(k_x, k_y + \frac{2\pi}{L_y})\rangle}, \nonumber
\end{align}
where the product over the momentum in both the denominator and numerator is the same, i.e., $k_a = \frac{2\pi}{L_a} n_a$ with $n_a \in \{1, 2, \cdots L_a\}$. For $L_a$ large, we can approximate
\begin{align}
\mathcal{Z}_c &\approx \frac{\prod_{k_x} \exp \Big[ i \int dk_y \mathcal{A}_y(k_x + \frac{\Phi_x}{L_x}, k_y)\Big]}{\prod_{ k_x} \exp \Big[ i \int dk_y \mathcal{A}_y(k_x, k_y)\Big]}\nonumber \\ 
&= \frac{\prod_{k_x} \exp \Big[ 2\pi i \mathcal{P}_y(k_x + \frac{\Phi_x}{L_x})\Big]}{\prod_{ k_x} \exp \Big[ 2\pi i \mathcal{P}_y(k_x)\Big]}, \nonumber
\end{align}
where $\mathcal{P}_y(k_x)$ is the $y$-directional polarization for the given momentum $k_x$. Finally Eq.~\eqref{Chern} reduces to:
 \begin{align}
\mathcal{Z}_c &= \prod_{k_x} \exp \Big[ 2\pi i \Big(\mathcal{P}_y \big(k_x + \frac{\Phi_x}{L_x}\big) -\mathcal{P}_y(k_x)\Big)\Big] \nonumber \\ 
&\approx \exp \Big[ i \Phi_x \int^{2 \pi}_0 dk_x \frac{d \mathcal{P}_y(k_x)}{d k_{x}}\Big] = \exp ( i \Phi_x C), 
\end{align}
where we have identified the change in the polarization $\mathcal{P}_y$ in $k_x$ as the Chern number~\cite{PhysRevLett.102.107603}. The generalization for the multiband cases is given in SI~\cite{suppl}.

We now use the many-body adiabatic pumping to show that Eq.~\eqref{Chern} must be related to the Hall conductivity $\sigma_{xy}$. To see this clearly, we use that the Resta operator $\hat{U}_{1,y}$ measures the many-body polarization. See SI for some details of $\hat{U}_{1,y}$~\cite{suppl}. Using the Resta operator, we find   
\begin{align}
\mathcal{Z}_c \propto \exp\Big[ 2\pi i \left(\text{P}_y (\Phi_x) - \text{P}_y (0) \right)\Big], \nonumber
\end{align}
where $\text{P}_y (\Phi_x)$ is the many-body polarization along $y$ for the ground state with the flux $\Phi_x$. For band insulators, $\text{P}_y (\Phi_x) = \sum_{k_x}\mathcal{P}_y \big( k_x + \frac{\Phi_x}{L_x} \big)$. When $\Phi_x$ is small, we find 
\begin{align}
\mathcal{Z}_c \approx \exp\Big[ 2\pi i \Phi_x \frac{d \text{P}_y }{d\Phi_x} \Big] =  \exp\Big[ 2\pi i \Phi_x \sigma_{xy} \Big] = \exp\Big(iC\Phi_x \Big) \nonumber 
\end{align}
where we noted that the polarization $\text{P}_y$ in the ground state can be induced by the adiabatic change in the flux $\Phi_x$, i.e., $\frac{d \text{P}_y }{d\Phi_x} = \frac{d \text{P}_y/dt }{d\Phi_x /dt} = \frac{J_y}{E_x} = \sigma_{xy}$. The last equality follows from $\sigma_{xy} = \frac{C}{2\pi}$. Here, we did not assume that the ground state is a band insulator and hence it suggests that Eq.\eqref{Chern} can be applied beyond band insulators. %, e.g., disordered and/or interacting cases without fractionalization. 

The final proof of Eq.~\eqref{Chern} in terms of the topological field theory interpretation is to show that it picks up the level $C$ of the Chern-Simons term, 
\begin{align}
S_{eff} = \frac{C}{4\pi} \int d\tau d^2 x ~\epsilon^{\mu\nu\lambda} A_\mu \partial_\nu A_\lambda,
\label{Chern3}
\end{align}
which is the effective field theory of the Chern insulator. Here we again do not assume that the ground state is a band insulator and so it shows that Eq.~\eqref{Chern} can be applied to any correlated insulator. We note that~\cite{suppl} 
\begin{align}
\langle \text{GS}[\Phi_x]| \hat{U}_{1,y} |\text{GS}[\Phi_x] \rangle \propto \exp\Big(iS_{eff}[A_\mu]\Big) = \exp \Big( i \Phi_x C \Big),  \nonumber
\end{align}
where $A_\mu$ is fixed by the twisted boundary condition $\Phi_x$ and the insertion of the unitary $\hat{U}_{1,y}$, i.e, $(A_0; A_x, A_y) = \big( \delta(\tau) \frac{2\pi}{L_y} y; \frac{\Phi_x}{L_x}, 0 \big)$. By inserting these gauge configurations into Eq.~\eqref{Chern3}, we indeed find that the level $C$ appears in the RHS of the above formula. We note that the same approach has been employed before~\cite{PhysRevB.100.245134}. 

%It is amusing to observe that Eq.~\eqref{Chern} unifies three distinct aspects of the Chern insulator, i.e., band index, adiabatic pumping, and the effective (topological) field theory, into a single expression. %\BK{Based on these three different analytic proofs, we provide numerical demonstration in Sec.~\ref{}, where consider three cases: free and translation-symmetric insulator, free but disordered insulator, and translation-symmetric but correlated insulators.} 

Before moving on, we comment on the relation of our formula Eq.~\eqref{Chern} with the previous related works. First, we note that a similar formula, which is essentially Eq.~\eqref{Chern} without the denominator, appeared in Ref. \onlinecite{PhysRevB.98.035151}. However, without the denominator, the formula suffers from an arbitrary phase correction to $\mathcal{Z}_c$, which lead the authors to keep track of the full adiabatic pumping of the flux insertion $\Phi_x \in [0, 2\pi)$ to calculate the Chern number. Second, there are expressions due to Niu-Thouless-Wu~\cite{PhysRevB.31.3372} and Hastings-Michalakis~\cite{hastings2015quantization}. The underlying idea of these expressions is to construct the effective Brillouin zone using the the twisted boundary conditions along the $x$- and $y$-direction following the idea of Ref.~\onlinecite{PhysRevB.31.3372}. For example, the one-plaquette formula has been introduced~\cite{PhysRevLett.122.146601} along these line of thoughts. However, the formula requires at least three different ground states, forming one plaquette, to compute the Chern number.

\textbf{3. Chiral Hinge Insulator}: We now present the many-body invariant for the $C_4 T$-symmetric chiral hinge insulator.
\begin{align}
\mathcal{Z}_h = \frac{\langle \text{GS}[\Phi_z]| \hat{U}_{2} |\text{GS}[\Phi_z] \rangle}{\langle \text{GS}[0]| \hat{U}_{2}|\text{GS}[0] \rangle} = \exp(i\Phi_z C_W)
\label{Hinge}
\end{align}
Here $|\text{GS}[\Phi_z ]\rangle$ is the ground state when the boundary condition along the $z$-axis is twisted by infinitesimal $\Phi_z$. We will see that $C_W$ is an integer labeling the quantized pumping of the quadrupole moment under the flux $\Phi_z$. In particular, when the insulator is the band insulator, $C_W$ will agree with the Wannier-sector Chern number.\cite{comment1}

%Interestingly, the above expression is written in real space coordinates and hence is complementary to the momentum-space invariant.\cite{PhysRevB.96.245115}

%For band insulators, $C_W$ agrees with the Wannier Chern number, defining the chiral hinge insulator. \BK{BK: I guess we don't have an explict proof of this statement.} \GYC{You are right --- we can erase this, perhaps.} $C_W$ effectively measures the adiabatic pumping of the quadrupole moment along $k_z$. Note that the above expression is written entirely in terms of real space coordinates and hence is complementary to the momentum-space invariant. %On the other hand, without refering to the band index, we can view $C_W$ as the coefficient of the effective field theory.  

First, we reduce Eq.~\eqref{Hinge} to a band index, i.e., the quantized pumping of the quadrupole moment along $k_z$. As before, we construct a many-body ground state as a Slater determinant of the occupied single-particle eigenstates. The single-particle states are given by the Bloch functions $|u_n (\bm{k}_\perp, k_z) \rangle$ and the plane waves $|\bm{k}_\perp, k_z \rangle$, i.e., $|\psi_n(\bm{k}_\perp, k_z) \rangle = |\bm{k}_\perp, k_z \rangle \otimes |u_n (\bm{k}_\perp, k_z) \rangle$ with $n$ labeling the filled bands and $\bm{k}_\perp$ being the momentum in the $xy$-plane. Then Eq.~\eqref{Hinge} reduces to
\begin{align}
\mathcal{Z}_h = \frac{\prod_{k_z} \prod_{\bm{k}_\perp, \bm{k}_\perp'} \mathcal{F}(\bm{k}'_\perp, \bm{k}_\perp, k_z +\frac{\Phi_z}{L_z})}{\prod_{k_z} \prod_{\bm{k}_\perp, \bm{k}_\perp'} \mathcal{F}(\bm{k}'_\perp, \bm{k}_\perp, k_z)}, 
\label{Hinge2}
\end{align}
where the product over the momentum $k_z$ is over the set $k_z \in \{\frac{2\pi}{L_z} n_z, n_z = 0, 1, \cdots L_z -1 \}$, and similarly for $\bm{k}$ and $\bm{k'}$. Here $\mathcal{F}$ is completely written out in terms of the Bloch functions: 
\begin{align}
 \mathcal{F}(\bm{k}'_\perp, \bm{k}_\perp, k_z +\frac{\Phi_z}{L_z}) = \langle \bm{k}_\perp', k_z + \frac{\Phi_z}{L_z}| e^{\frac{i2\pi xy}{L_x L_y}}|\bm{k}_\perp, k_z + \frac{\Phi_z}{L_z} \rangle \nonumber \\ \prod_{n,m} \langle u_n \left(\bm{k}_\perp', k_z + \frac{\Phi_z}{L_z}\right)| u_m \left(\bm{k}_\perp, k_z + \frac{\Phi_z}{L_z} \right)\rangle, \nonumber 
\end{align}
where the product is over occupied bands. We identify
\begin{align}
\prod_{\bm{k}_\perp, \bm{k}_\perp'} \mathcal{F} \Big(\bm{k}'_\perp, \bm{k}_\perp, k_z +\frac{\Phi_z}{L_z} \Big) = \exp \left[2\pi i \mathcal{Q}_{xy}\Big(k_z + \frac{\Phi_z}{L_z} \Big)\right], \nonumber 
\end{align} 
where $\mathcal{Q}_{xy}(k_z)$ is the quadrupole moment of an effective two-dimensional model when $k_z$ is viewed as an adiabatic parameter.\cite{PhysRevB.100.245134, PhysRevB.100.245135} We assume the nonzero Wannier gap, which is required for defining quadrupole moment.\cite{benalcazar2017quantized,PhysRevB.100.245134, PhysRevB.100.245135} (See Refs.[\onlinecite{PhysRevB.100.245133, PhysRevB.100.245134, PhysRevB.100.245135}] for the cases without the Wannier gap.)  Now, Eq.~\eqref{Hinge2} for infinitesimal $\Phi_z$ becomes  
\begin{align}
\mathcal{Z}_h &= \prod_{k_z} \exp\Big[2\pi i \left(\mathcal{Q}_{xy} \Big(k_z + \frac{\Phi_z}{L_z} \Big) -\mathcal{Q}_{xy}(k_z) \right)\Big] \nonumber\\ 
& \approx \exp \Big[ i \Phi_z \int^{2 \pi}_{0} dk_z \frac{d \mathcal{Q}_{xy}}{dk_z} \Big] =   \exp \Big[i \Phi_z C_W \Big],  \nonumber 
\end{align} 
i.e., $\mathcal{Z}_h$ measures the chiral pumping of quadrupole moment $\mathcal{Q}_{xy} (k_z)$ along $k_z$, as advertised.

Remarkably, our formula Eq.~\eqref{Hinge} is consistent with the effective field theory of the chiral hinge insulator proposed in Ref.~[\onlinecite{you2019multipolar}] 
\begin{align}
S_{eff}^h = \frac{C_W}{4\pi} \int d\tau d^3 x ~ A_z \left(\partial_t \partial_y A_x + \partial_t \partial_x A_y -2 \partial_x \partial_y A_0 \right).  \nonumber 
\end{align}
To relate Eq.~\eqref{Hinge} to this effective theory, we note~\cite{suppl} 
\begin{align}
\langle \text{GS}| \hat{U}_{2} |\text{GS} \rangle \propto \text{Tr} \Big[ e^{-\beta \hat{H}_0} \hat{U}_2 \Big]_{\beta \rightarrow \infty} \propto \exp \Big(iS_{eff}^h[A_\mu]\Big), \nonumber
\end{align}
where $A_\mu$ is fixed by the twisted boundary condition $\Phi_z$ and the insertion of the unitary $\hat{U}_{2}$, i.e, $(A_0; A_x, A_y, A_z) = (\delta(\tau) \frac{2\pi xy}{L_y L_x}; 0, 0, \frac{\Phi_z}{L_x})$. On inserting this gauge field configuration to the effective theory, we find Eq.~\eqref{Hinge}. Note that the effective topological field theory is expected to remain valid for interacting and/or disordered cases. Hence, we expect the same for Eq.\eqref{Hinge}.  

Finally, we discuss the physics hidden behind Eq.~\eqref{Hinge}. This highlights the relation of our formula Eq.~\eqref{Hinge} with the Wannier-sector Chern number.\cite{PhysRevB.100.245134, PhysRevB.100.245135} We first note that the ground state overlaps in Eq.~\eqref{Hinge} give the bulk quadrupole moments along the $xy$-direction of the ground states. 
\begin{align}
\mathcal{Z}_h \propto \exp\Big[ 2\pi i \left(\text{Q}_{xy} (\Phi_z) - \text{Q}_{xy} (0) \right)\Big], \nonumber
\end{align}  
where $\text{Q}_{xy}(\Phi_z)$ is the $xy$-plane quadrupole moment of the ground state with the flux $\Phi_z$. Assuming that $\text{Q}_{xy}$ is a smooth function of $\Phi_z$, we find: 
\begin{align}
\mathcal{Z}_h \propto e^{2\pi i  \Phi_z \frac{d\text{Q}_{xy}}{d\Phi_z}} \to e^{2\pi i  \Phi_z \frac{d\text{Q}_{xy}/dt}{d\Phi_z/dt}} = e^{2\pi i \Phi_z \frac{J_{xy}^{\text{quad}}}{E_{z}}}. \label{Hinge-Streda}
\end{align}
Here, we imagined to change the flux $\Phi_z$ adiabatically in time, which will induce the electric field along the $z$-direction on the surface. The change in quadrupole moment is given by the surface current $J_{\text{quad}}$ perpendicular to the $z$-direction. Now, by insisting that the last expression on Eq.~\eqref{Hinge-Streda} to be equivalent to Eq.~\eqref{Hinge}, we find 
\begin{align}
\sigma_{xy}^{\text{quad}} = \frac{J_{xy}^{\text{quad}}}{E_{z}} = \frac{C_W}{2\pi}. \nonumber
\end{align}
This is precisely what one expects from the nested Wilson loop picture\cite{PhysRevB.96.245115}, and provides additional supporting evidence for our formula. We complete the comparison by performing another transformation on Eq.~\eqref{Hinge-Streda}. Let us rewrite $J_{xy}^{\text{quad}}$ as the variation of the quadrupolar surface orbital magnetization $M_{xy}^{\text{quad}}$, i.e, $J_{xy}^{\text{quad}}= \frac{d M_{xy}^{\text{quad}}}{dz}$. Similarly, we rewrite $E_z = \frac{d \mu^s}{dz}$, where $\mu^s$ is the surface chemical potential. Then, we find  
\begin{align}
\mathcal{Z}_h =  e^{2\pi i \Phi_z \frac{J_{xy}^{\text{quad}}}{E_{z}}} \to e^{2\pi i \Phi_z \frac{dM_{xy}^{\text{quad}}}{d\mu^s}}. \nonumber 
\end{align}
Enforcing the last expression to agree with Eq.~\eqref{Hinge-Streda}, we find 
\begin{align}
\frac{dM_{xy}^{\text{quad}}}{d\mu^s} = \frac{C_W}{2\pi}. \nonumber
\end{align}
This is again consistent with the nested Wilson loop picture\cite{PhysRevB.96.245115} and the orbital magnetization calculation\cite{PhysRevB.74.024408}. Hence, from Eq.~\eqref{Hinge}, we complete our derivation of the quadrupolar ``Streda formula":  
\begin{align}
\frac{dQ_{xy}}{d\Phi_z} = \frac{J_{xy}^{\text{quad}}}{E_z} = \frac{dM_{xy}^{\text{quad}}}{d\mu^s} = \frac{C_W}{2\pi}. \nonumber
\end{align}
This is consistent with the effective theory.\cite{you2019multipolar} 

\textbf{4. Numerical demonstrations}: 
In this section, we provide numerical confirmation of our many-body invariants for the two classes of the chiral topological insulators that we discussed above. We demonstrate the validity of our invariants for various models including non-interacting translation-symmetric insulators, non-interacting but without translation symmetry due to weak disorder, and correlated insulators. In all cases, our many-body invariants faithfully measure the bulk topological invariants. %We also emphasize that our method provides an efficient way of detecting the underlying topology as the invariants use only 2 ground states.

\begin{figure}[t]
\centering\includegraphics[width=0.45\textwidth]{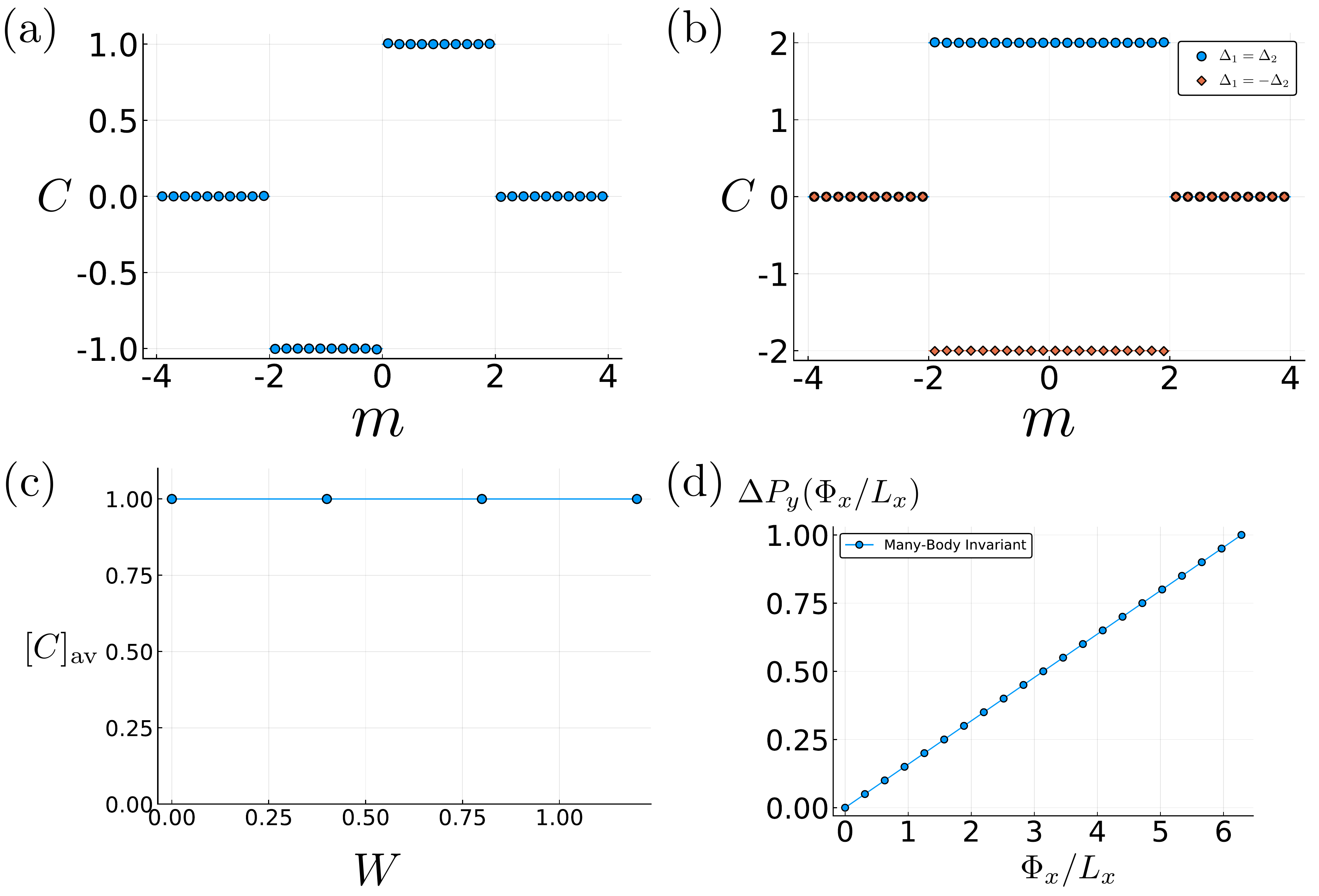}
\caption{The Chern numbers from Eq.~\eqref{Chern} for (a) $\hat{H}_\textrm{Ch}^{(1)}$ with $(t, \Delta) = (1.0, 1.0)$ and (b) $\hat{H}_\textrm{Ch}^{(2)}$ with $(t, \Delta_1, \Delta_2) = (1.0, 1.0, \pm 1.0)$ as a function of $m$. (c) The disorder averaged Chern number as a function of the mass disorder strength $W$. (d) Change in the phase factor of $\mathcal{Z}_c$ as a function of $\Phi_x/L_x$ for $\hat{H}_\textrm{Ch}^{(1)} + \hat{H}_U$ at $U=1$. }
\label{Fig:Cherns}
\end{figure}

We begin with two non-interacting tight-binding models of Chern insulators~\cite{PhysRevB.98.235160} on a square lattice:
\begin{align}
\hat{H}_\textrm{Ch}^{(1)} = \sum_{\bm k} c_{\bm k}^\dagger \big[ &(m - t \cos(k_x) - t \cos(k_y) ) \sigma_z \nonumber \\
&+ \Delta \sin(k_x) \sigma_x + \Delta \sin(k_y) \sigma_y  \big] c_{\bm k}
\label{Eq:Chern1}
\end{align}
and 
\begin{align}
\hat{H}_\textrm{Ch}^{(2)} = \sum_{\bm k} c_{\bm k}^\dagger \big[ &(m - t\cos(k_x) - t \cos(k_y)) \sigma_z \nonumber \\
& + \Delta_1 (\cos(k_x) - \cos(k_y) ) \sigma_y \nonumber \\
&+ \Delta_2 \sin(k_x) \sin(k_y) \sigma_x \big] c_{\bm k}
\label{Eq:Chern2}
\end{align}
where $c_{\bm k}^\dagger = (c_{\bm{k}, A}^\dagger, c_{\bm{k}, B}^\dagger)$. By tuning the parameter $m$ while keeping $t$ and $\Delta$ ($\{\Delta_1, \Delta_2 \}$) fixed, the half-filled ground state of $\hat{H}_\textrm{Ch}^{(1)}$ ($\hat{H}_\textrm{Ch}^{(2)}$) realizes the quantum phase transitions between the $C = \pm 1$ ($C = \pm 2$) Chen insulator to the $C=0$ trivial insulator. In FIG.~\ref{Fig:Cherns} (a) and (b), we confirm that our many-body invariant Eq.~\eqref{Chern} well reproduces the expected phase diagrams of $\hat{H}_\textrm{Ch}^{(1)}$ and $\hat{H}_\textrm{Ch}^{(2)}$. 

To further confirm the validity of our invariants, we consider a non-interacting model with onsite disorder, by replacing the uniform mass term $\sum_{\bm k} m c_{\bm k}^\dagger \sigma_z c_{\bm k}$ in Eq.~\eqref{Eq:Chern1} to position dependent random mass term $\sum_{\bm r} m_{\bm r} c_{\bm r}^\dagger \sigma_z c_{\bm r}$ where $c_{\bm r}^\dagger = (c_{\bm{r}, A}^\dagger, c_{\bm{r}, B}^\dagger)$ and $m_{\bm r} \in [1.0, 1.0+W]$. In FIG.~\ref{Fig:Cherns} (c), the disorder averaged Chern number $[C]_\textrm{av}$, which is the average of the Chern number Eq.~\eqref{Chern} over 1,000 disorder realizations, shows stable behavior as a function of disorder strength $W$. This shows the stability of our formula against weak disorders. We leave strong disorder behaviors, e.g., disorder-driven topological phase transitions, of Eq.\eqref{Chern} to the future work.

Finally, we consider an interacting model $\hat{H} = \hat{H}_\textrm{Ch}^{(1)} + \hat{H}_U$, where $\hat{H}_\textrm{Ch}^{(1)}$ is given by Eq.~\eqref{Eq:Chern1}. $\hat{H}_U = U \sum_{\bm r} c_{\bm{r}, A}^\dagger c_{\bm{r}, A} c_{\bm{r},B}^\dagger c_{\bm{r}, B}$ is the onsite Hubbard interaction. We perform an infinite density matrix renormalization group (iDMRG) simulation~\cite{PhysRevLett.69.2863, SCHOLLWOCK201196} on an infinite cylinder with finite circumference, where the $x$-axis is the infinite direction parallel to the cylindrical axis and the $y$-direction contains finite $N_y$ sites. From the infinite matrix product state (iMPS) generalization of our many-body invariant~\cite{suppl}, we can extract the Chern number using two iMPS ground states---one without any flux insertion and the other with an infinitesimal flux inserted along the $x$-direction. 

Our method bears some similarity with the usual method for measuring the Chern number in the iDMRG simulations~\cite{zaletel2014flux}. In fact, both methods measure the identical observables when $N_y$ is sufficiently large. However, when $N_y$ is finite and small, which is the case for the usual iDMRG simulations on quasi-1d systems, two methods measure the Chern number in a slightly different way. The usual method~\cite{zaletel2014flux} measures the polarization along the $x$-direction, the infinite direction, {\it exactly}, but twisting the boundary condition {\it adiabatically} by threading the flux $\Phi_y$ from $0$ to $2 \pi$ along the $y$-direction, the finite circumference direction. Since the allowed momentum along the $y$-direction is highly restricted due to the smallness of $N_y$, it is important to keep track of the winding of the $x$-polarization with respect to the full adiabatic changes of $\Phi_y$. On the other hand, our method measures the polarization along the $y$-direction {\it approximately} using $\hat{U}_{1,y}$ operator, but only requires an {\it infinitesimal} gauge flux along the $x$-direction, which is the infinite direction. Since the momentum along the infinite direction is fully available, it is sufficient to consider the states with $0$ and an infinitesimal flux. In the iDMRG simulation, we set $(t, m, r; U) = (1.0,1.0,1.0;1.0)$, $N_y = 4$, and the bond dimension $\chi = 100$. In FIG.~\ref{Fig:Cherns}, we see that the polarization along the $y$-direction, which is computed from the many-body invariant, shows perfect linear behavior as a function of the flux $\Phi_x$ along the $x$-direction, thereby indicating that an infinitesimal flux insertion is enough for computing the Chern number.

\begin{figure}[t]
\centering\includegraphics[width=0.5\textwidth]{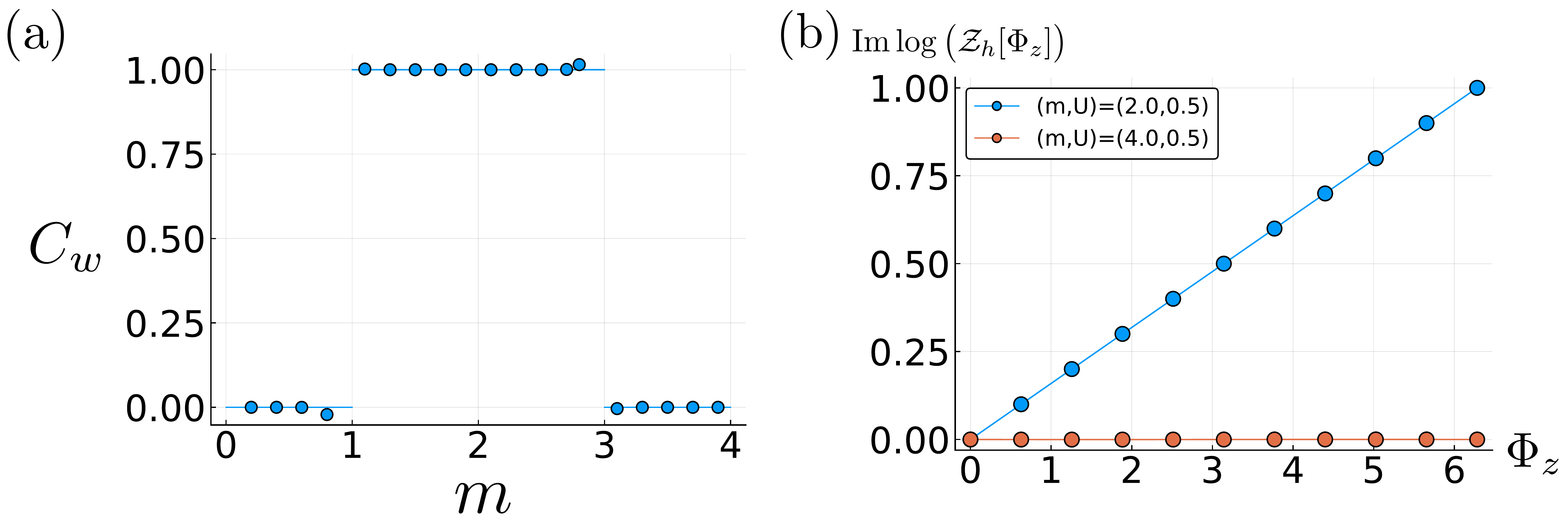}
\caption{(a) $C_W$ from Eq.\eqref{Hinge} for the half-filled ground state of Eq.\eqref{Eq:Hinge-TB} with $(t, \Delta_1, \Delta_2) = (1.0, 1.0, 1.0)$, system size $(L_x, L_y, L_z) = (20, 20, 40)$, and $\Phi_z = 0.01$. (b) Change in the phase factor of $\mathcal{Z}_h (\Phi_z)$ as a function of $\Phi_z$. The ground states are obtained within the Hartree-Fock approximation for $(m,U)=(2.0,0.5)$ and $(4.0, 0.5)$. The slope gives $C_W$. We see perfect linear behaviors in both cases.}
\label{Fig:Hinges}
\end{figure}

Finally, we confirm the many-body invariant Eq.~\eqref{Hinge} for the chiral hinge insulators. We first consider the following tight-binding model~\cite{schindler2018higher}:
\begin{align}
\label{Eq:Hinge-TB}
&\hat{H}_{\rm Hinge} =\sum_{\bm k} c_{\bm k}^\dagger \big[(m - t \sum_{i=x,y,z} \cos(k_i) ) \tau_z \sigma_0  \\
&+ \Delta_1 \sum_{i=x,y,z} \sin(k_i) \tau_x \sigma_i + \Delta_2 (\cos(k_x) - \cos(k_y)) \tau_y \sigma_0 \big] c_{\bm k}, \nonumber
\end{align}
where $c_{\bm k}=(c_{\bm{k}, A, \uparrow}, c_{\bm{k}, A, \downarrow}, c_{\bm{k}, B, \uparrow}, c_{\bm{k}, B, \downarrow})^T$. In FIG.~\ref{Fig:Hinges} (a), we use the many-body invariant Eq.~\eqref{Hinge} to compute $C_W$ as a function of the mass parameter $m$ with fixed $(t, \Delta_1, \Delta_2) = (1.0, 1.0, 1.0)$. We see that the many-body invariant Eq.~\eqref{Hinge} reproduces the phase diagram with the correct $C_W$.

For a non-trivial test for our invariant Eq.~\eqref{Hinge}, we add an onsite Hubbard interaction $U \sum_{\bm r, a=A,B} \hat{n}_{\bm{r}, a, \uparrow} \hat{n}_{\bm{r}, a, \downarrow}$ to Eq.~\eqref{Eq:Hinge-TB}. To include the interaction effects, we employ the self-consistent Hartree-Fock method which variationally finds the single Slater determinant wavefunction minimizing the energy of the interacting Hamiltonian. We turn on small yet finite $U$ where the ground state is adiabatically connected to the ground state with $U=0$. We consider the full $2\pi$ flux insertion along the $z$-direction and see how the many-body invariant Eq.~\eqref{Hinge} changes. In FIG.~\ref{Fig:Hinges} (b), we observe that the slope of $\textrm{Im} \log [ \mathcal{Z}_h (\Phi_z) ]$ equals $C_W$ as expected. 

We also have investigated Eq.~\eqref{Hinge} for the chiral hinge insulators, whose chiral modes are purely due to the boundary Chern bands and not due to the pumping of bulk quadrupole moments. For these cases, we expect them to be identified with $C_W =0$ in Eq.~\eqref{Hinge}. We will report the detection of this boundary Chern insulators using Eq.~\eqref{Hinge} as well as the applicability of $\hat{U}_2$ in measuring various boundary-only topology elsewhere.~\cite{boundary_Chern}

%\begin{figure}[t]
%\centering\includegraphics[width=0.40\textwidth]{Figure-Hinge.pdf}
%\caption{Wannier Chern number from the many-body invariant for Hinge insulator Eq.~\eqref{Hinge}. We see that the many-body invariant reproduce the expected phase diagram, especially phase boundaries.}
%\label{Fig:Hinge}
%\end{figure}

%\begin{figure}[t]
%\centering\includegraphics[width=0.40\textwidth]{Figure-Hinge-HF.pdf}
%\caption{The winding of $\hat{U_2}$ as a function of $z$-flux $\Phi_z$ in the case of interacting Hinge insulator. We set $(t, \Delta_1, \Delta_2) = (1.0,1.0,1.0)$ with system size $L = 10$ and add on-site Hubbard interaction having strength $U$. The ground state is constructed from self-consistent Hartree-Fock method for topological Hinge insulator $(m, U) = (2.0, 0.5)$ and trivial Hinge insulator $(m, U) = (4.0, 0.5)$. We see that our many-body invariant gives perfect linear line over the whole flux $\Phi_z$. \GYC{Please merge the figures for the cleaner outlook of the page.}}
%\label{Fig:Hinge-HF}
%\end{figure}

\textbf{5. Conclusions}: 
We have provided many-body invariants for 2d Chern and 3d chiral hinge insulators. Our many-body invariants have natural interpretations in three different theoretical approaches: many-body adiabatic pumping, topological field theory, and free-fermion band indices. Our invariants provide not only the theoretical framework to understand various chiral insulators, but also efficient ways of computing topological numbers as the invariants require only two ground states. We have tested our formula for various non-interacting models with translation symmetry, non-interacting but without translational symmetry, and interacting models. In all cases, we see that our invariants detect the ground state topology correctly and successfully map out the full phase diagram as well. 

%Finally it is not difficult to anticipate the generalization of our invariants to other topological insulators. 

%It would be interesting to find the connections between our invariants and other methods of measuring topological invariants such as the Chern number in terms of projectors supported on regions on real space~\cite{kitaev2006anyons} and recently proposed band index for chiral hinge insulator~\cite{PhysRevLett.124.036401}, which we leave as a future work.

\acknowledgements
We thank Barry Bradlyn, Jong Yeon Lee, Masaki Oshikawa, and Aaron Szasz for helpful discussion. BK is supported by KIAS individual Grant PG069401 at Korea Institute for Advanced Study. This work of BK is supported by the Center for Advanced Computation at Korea Institute for Advanced Study. This work of GYC was supported by the National Research Foundation of Korea(NRF) grant funded by the Korea government(MSIT) (No. 2020R1C1C1006048). GYC also acknowledge the hospitality of Korea Institute for Advanced Study, where some part of this work has been done.

\bibliography{Many-Body}

\begin{thebibliography}{33}
\bibitem [{com()}]{comment1}%
  \BibitemOpen
  \href@noop {} {}\bibinfo {note} {Here $C_W$ counts the quantized pumping of
  the ``physical" quadrupole moment, which is shown to be captured by the
  Wannier Chern number in [\onlinecite{PhysRevB.96.245115}].}\BibitemShut
  {Stop}%
\end{thebibliography}

\end{document}

\title{Supplemental Information for ``Many-Body Invariants for Chern and Chiral Hinge Insulators"}

\author{Byungmin Kang}
\thanks{Electronic Address: bkang119@kias.re.kr}
\affiliation{School of Physics, Korea Institute for Advanced Study, Seoul 02455, Korea}

\author{Wonjun Lee}
\thanks{Electronic Address: wonjun1998@postech.ac.kr}
\affiliation{Department of Physics, Pohang University of Science and Technology (POSTECH), Pohang 37673, Republic of Korea}

\author{Gil Young Cho}
\thanks{Electronic Address: gilyoungcho@postech.ac.kr}
\affiliation{Department of Physics, Pohang University of Science and Technology (POSTECH), Pohang 37673, Republic of Korea}

\date{\today}
\maketitle
\tableofcontents
\appendix

\section{On the Unitaries $\hat{U}_{a=1,2}$}
In this section, we briefly review the necessary information about the unitary operators $\hat{U}_{a=1,2}$ which are extensively used in the main text. We ask readers to refer to two papers, one~\cite{PhysRevLett.80.1800} by Resta and the other~\cite{PhysRevB.100.245134, PhysRevB.100.245135} by Kang, Shiozaki and Cho, or Wheeler, Wagner, and Hughes for more details. In summary, the unitary operators $\hat{U}_{a=1,2}$ are designed to measure the polarization and quadrupole moments of many-body ground states. 

\subsection{Many-Body Polarization}
The unitary $\hat{U}_1$ was used to measure the polarization. The statement in the original paper~\cite{PhysRevLett.80.1800} is that the polarization of an 1d insulator, which could be free or correlated, can be purely diagnosed from its ground state defined on a torus with the length $L$ under the periodic boundary condition, i.e., each lattice site is labeled as $x \in \mathbb{Z}$ such that $x \sim x+ L$. Then, we consider a unitary $\hat{U}_{1}$:
\begin{align}
\hat{U}_1 = \exp \Big( \frac{2\pi i}{L} \sum_{x=1}^{L} x \hat{n}_x \Big),
\end{align} 
where $\hat{n}_x$ is the number operator of the electrically-charged particles at the site $x$. Then, we can compute the many-body polarization in terms of the many-body ground state $\vert \textrm{GS} \rangle$ on the torus 
\begin{align}
\langle \textrm{GS}  \vert \hat{U}_1 \vert \textrm{GS} \rangle = |\langle \hat{U}_1 \rangle| \exp (2\pi i \text{P}), 
\end{align}
where $\text{P}$ is the polarization. Note that in the case of the band insulators, the polarization P appeared in the above definition can be identified with the polarization defined in terms of the Berry connection as $L \to \infty$.~\cite{PhysRevLett.80.1800} 

\subsection{Many-Body Quadrupole}
The quadrupole moment of 2d insulator can be similarly measured as the many-body polarization. We start with the many-body ground state $\vert \textrm{GS} \rangle$ of a 2d insulator on the torus of the length $L_x \times L_y$ with the periodic boundary conditions along both directions, i.e., $(x,y) \in \mathbb{Z}^2$ such that $(x,y) \sim (x+L_x, y) \sim (x, y+L_y) \sim (x+L_x, y+L_y)$. Next, we consider a unitary $\hat{U}_2$:
\begin{align}
\hat{U}_2 = \exp \Big( \frac{2\pi i}{L_x L_y} \sum_{x=1}^{L_x} \sum_{y=1}^{L_y} xy \hat{n}_{x,y} \Big),
\end{align} 
where $\hat{n}_{x,y}$ is the number operator of the electrically-charged particles at the site $(x,y)$. Following Refs.~\onlinecite{PhysRevB.100.245134, PhysRevB.100.245135}, the bulk quadrupole moment of the insulator can be measured as
\begin{align}
\langle \textrm{GS} \vert \hat{U}_2 \vert \textrm{GS} \rangle = |\langle \hat{U}_2 \rangle| \exp (2\pi i \text{Q}_{xy}), 
\end{align}
where $\text{Q}_{xy}$ is the (physical) quadrupole moment of the insulator. Note that the polarization $\text{Q}_{xy}$ as defined above agrees with the quadrupole moment defined in terms of the Bloch states and associated nested Wilson loop index, when the $C_4$ symmetry or the two anti-commuting mirror symmetries $R_x \times R_y$ exist in the system.~\cite{benalcazar2017quantized, PhysRevB.100.245134, PhysRevB.100.245135} Currently, the many-body expectation value and the physical quadrupole moments seem to agree with each other for the band insulators with the Wannier gaps\cite{PhysRevB.100.245134, PhysRevB.100.245135}; it is an open problem if one can define the bulk quadrupole moments for the cases without the Wannier gaps. The detailed discussions on applying $\hat{U}_2$ on the cases without Wannier gaps can be found in Refs. [\onlinecite{PhysRevB.100.245134, PhysRevB.100.245135,PhysRevB.100.245133,boundary_Chern}].

\section{Reduction to Band Chern Number}
In this section, we show how the many-body invariant for Chern number is reduced to the band Chern number for the ground state of band insulators. We assume translational symemtries along the $x$- and $y$-direction and fix the convention for the Bloch states as
\begin{equation}
c_{(k_x, k_y), \alpha}^\dagger = \frac{1}{\sqrt{L_x L_y}} \sum_{x=1}^{L_x} \sum_{y=1}^{L_y} e^{-i (k_x x + k_y y)} c_{(x,y), \alpha}^\dagger,
\end{equation}
where $L_x$ and $L_y$ are system sizes along the $x$- and $y$-direction and $\alpha$ labels the orbtial degrees of freedom. Using the bloch states, a non-interacting tight-binding hamiltonian can be written as
\begin{equation}
\hat{H} = \sum_{k_x, k_y} \sum_{\alpha, \beta=1}^{N_\textrm{orb}} h_{\alpha, \beta}(k_x, k_y) \hat{c}_{(k_x, k_y), \alpha}^\dagger \hat{c}_{(k_x, k_y), \beta},
\label{eq:TB-ham}
\end{equation}
where $N_\textrm{orb}$ is the number of orbitals in the unit-cell and the momentum summation is over $k_x \in \{0, \frac{2\pi}{L_x}, \cdots, \frac{2\pi (L_x - 1)}{L_x} \}$ and $k_y \in \{ 0, \frac{2\pi}{L_y}, \cdots, \frac{2\pi (L_y - 1)}{L_y} \}$, i.e., periodic boundary condition along the $x$- and $y$-direction. By diagonalizing $h_{\alpha, \beta} (k_x, k_y)$, one gets single-particle eigenstates 
\begin{equation}
\hat{\gamma}_{n, (k_x, k_y)}^\dagger = \sum_{\alpha=1}^{N_\textrm{orb}} u_{n, \alpha}(k_x, k_y) \hat{c}_{(k_x, k_y), \alpha}^\dagger ,
\end{equation}
where $n \in \{1, 2, \cdots, N_\textrm{orb}\}$ is the band index. 

Let us consider a band insulator where the many-body ground state can be obtained by completely fill the lowest $N_\textrm{occ}$ bands:
\begin{equation}
\vert \textrm{GS} \rangle = \Big( \prod_{n, k_x, k_y} \hat{\gamma}_{n, (k_x, k_y)}^\dagger \Big) \vert \textrm{vac} \rangle,
\label{eq:GS-TB-ham}
\end{equation}
where $\vert \textrm{vac} \rangle$ is the vacuum state and we fix, once it for all, the ordering of the product of single-particle creation operators.

We now carefully analyze the gauging of the $U(1)$-symmetry in the case of our non-interacting tight-binding hamiltonian Eq.~\eqref{eq:TB-ham}. To this end, we introduce a constant gauge field $(A_x, A_y) = (\Phi_x/L_x, \Phi_y/L_y)$ which leads us to the following changes in the Hamiltonian:
\begin{equation}
\hat{H} (\Phi_x, \Phi_y) = \sum_{k_x, k_y} \sum_{\alpha, \beta=1}^{N_\textrm{orb}} h_{\alpha, \beta}\Big(k_x + \frac{\Phi_x}{L_x}, k_y + \frac{\Phi_y}{L_y} \Big) \hat{c}_{(k_x, k_y), \alpha}^\dagger \hat{c}_{(k_x, k_y), \beta},
\end{equation}
where the summation of $(k_x, k_y)$ is over the same set of (discrete) momenta as in Eq.~\eqref{eq:TB-ham}. Similarly, the ground state of $\hat{H} (\Phi_x, \Phi_y)$ can be written as
\begin{equation}
\vert \textrm{GS} (\Phi_x, \Phi_y) \rangle = \Big( \prod_{n, k_x, k_y} \hat{\gamma}_{n, \big(k_x + \frac{\Phi_x}{L_x}, k_y + \frac{\Phi_y}{L_y} \big)}^\dagger \Big) \vert \textrm{vac} \rangle,
\end{equation}
where we use the same ordering of the product as the one used in Eq.~\eqref{eq:GS-TB-ham} and
\begin{equation}
\hat{\gamma}_{n, \big(k_x + \frac{\Phi_x}{L_x}, k_y + \frac{\Phi_y}{L_y} \big)}^\dagger = \sum_{\alpha=1}^{N_\textrm{orb}} u_{n, \alpha}\big(k_x + \frac{\Phi_x}{L_x}, k_y + \frac{\Phi_y}{L_y} \big) \hat{c}_{(k_x, k_y), \alpha}^\dagger 
\end{equation}
with $u_{n, \alpha}\big(k_x + \frac{\Phi_x}{L_x}, k_y + \frac{\Phi_y}{L_y} \big)$ being the eigenstate of the Bloch Hamiltonian $h_{\alpha, \beta}\Big(k_x + \frac{\Phi_x}{L_x}, k_y + \frac{\Phi_y}{L_y} \Big)$. Note that the creation operator part $\hat{c}_{(k_x, k_y), \alpha}^\dagger$ remains the same while its amplitude changes from $u_{n, \alpha} (k_x, k_y)$ to $u_{n, \alpha}\big(k_x + \frac{\Phi_x}{L_x}, k_y + \frac{\Phi_y}{L_y} \big)$.

Before proceed, let us remark the relation between $U(1)$-gauge flux $\Phi_y$ and the many-body operator $\hat{U}_{1;y} = \exp \big( \frac{2\pi i}{L_y} \sum_{x,y} y \hat{n}_{(x,y)}\big)$. For an arbitrary $\Phi_y$, the Hamiltonian is not gauge equivalent to the original Hamiltonian with $\Phi_y = 0$. But when $\Phi_y = 2\pi$ (or a multiple of $2\pi$), the Hamiltonian is gauge-equvalent to the original Hamiltonian via $\hat{U}_{1;y}$: 
\begin{equation}
\hat{H}(\Phi_y = 2\pi) = \hat{U}_{1;y} \hat{H}(\Phi_y = 0) \hat{U}_{1;y}^\dagger ,
\end{equation}
which also holds in the case of fully interacting many-body Hamiltonian with $U(1)$-symmetry (charge conservation). The Hamiltonian Eq.~\eqref{eq:TB-ham} does follow the same transformation, which can be directly seen by using
\begin{align}
\hat{U}_{1;y} \big( \hat{c}_{(k_x, k_y), \alpha}^\dagger \big) \hat{U}_{1;y}^\dagger &= \frac{1}{\sqrt{L_x L_y}} \sum_{x'=1}^{L_x} \sum_{y'=1}^{L_y} e^{-i(k_x x' + k_y y')} \hat{U}_{1;y} \big( \hat{c}_{(x',y'), \alpha}^\dagger \big) \hat{U}_{1;y}^\dagger \nonumber \\
&= \frac{1}{\sqrt{L_x L_y}} \sum_{x'=1}^{L_x} \sum_{y'=1}^{L_y} e^{-i(k_x x' + k_y y')} e^{\frac{2\pi i}{L_y} y'} \hat{c}_{(x',y'), \alpha}^\dagger \nonumber \\
&= \hat{c}_{\big(k_x, k_y - \frac{2\pi}{L_y} \big), \alpha}^\dagger
\end{align}
and relabel the dummy variable $k_y$ in the summation in Eq.~\eqref{eq:TB-ham}.

Our many-body index for Chern number amounts to computing the expectation value of $\hat{U}_{1;y}$ with respect to the ground states $\vert \textrm{GS} (0, 0) \rangle$ and $\vert \textrm{GS} (\Phi_x, 0) \rangle$ for an infinitesimal $\Phi_x$. Using the fact that $\hat{U}_{1;y} \vert \textrm{vac} \rangle = \vert \textrm{vac} \rangle$ and 
\begin{equation}
\hat{U}_{1;y} \hat{\gamma}_{n, \big(k_x + \frac{\Phi_x}{L_x}, k_y \big)}^\dagger \hat{U}_{1;y}^\dagger = \sum_{\alpha=1}^{N_\textrm{orb}} u_{n, \alpha}\big(k_x + \frac{\Phi_x}{L_x}, k_y \big) \hat{c}_{(k_x, k_y - \frac{2\pi}{L_y}), \alpha}^\dagger ,
\end{equation}
$\hat{U}_{1;y} \vert \textrm{GS} (\Phi_x, 0) \rangle$ can be written as
\begin{equation}
\hat{U}_{1;y} \vert \textrm{GS} (\Phi_x, 0) \rangle = \pm \prod_{n, k_x, k_y} \bigg[  \sum_{\alpha=1}^{N_\textrm{orb}} u_{n, \alpha}\big(k_x + \frac{\Phi_x}{L_x}, k_y + \frac{2\pi}{L_y} \big) \hat{c}_{(k_x, k_y), \alpha}^\dagger \bigg] \vert \textrm{vac} \rangle,
\label{Eq:U1y-GS-Phi_x}
\end{equation}
where $\pm$ is a possible sign factor which appears when re-arranging fermionic operators in order to follow the ordering of the product in Eq.~\eqref{eq:GS-TB-ham}. The single-particle creation operators in $\hat{U}_{1;y} \vert \textrm{GS} (\Phi_x, 0) \rangle$ still caries momentum quantum number, which simplifies the inner product between $\vert \textrm{GS} (\Phi_x, 0) \rangle$ and $\hat{U}_{1;y} \vert \textrm{GS} (\Phi_x, 0) \rangle$:
\begin{equation}
\langle \textrm{GS} (\Phi_x, 0) \vert \hat{U}_{1;y} \vert \textrm{GS} (\Phi_x, 0) \rangle = \pm \prod_{k_x, k_y} \det \bigg[\Big\langle u_n (k_x + \frac{\Phi_x}{L_x}, k_y) \Big\vert u_m\big(k_x + \frac{\Phi_x}{L_x}, k_y + \frac{2\pi}{L_y} \big) \Big\rangle \bigg],
\label{Eq:GS-overlap}
\end{equation}
where $\pm$ is a possible sign factor, in addition to the sign factor in Eq.~\eqref{Eq:U1y-GS-Phi_x}, which arises when grouping single-particle operators with the same momentum quantum number. For each $(k_x, k_y)$, we have $N_\textrm{occ}$-by-$N_\textrm{occ}$ matrix whose $(n,m)$th entry is given by $\Big\langle u_n (k_x + \frac{\Phi_x}{L_x}, k_y) \Big\vert u_m\big(k_x + \frac{\Phi_x}{L_x}, k_y + \frac{2\pi}{L_y} \big) \Big\rangle = \sum_{\alpha=1}^{N_\textrm{orb}} u_{n, \alpha}^* (k_x + \frac{\Phi_x}{L_x}, k_y) u_{m, \alpha} \big(k_x + \frac{\Phi_x}{L_x}, k_y + \frac{2\pi}{L_y} \big)$. The determinant in the RHS of Eq.~\eqref{Eq:GS-overlap} denotes the determinant of this $N_\textrm{occ}$-by-$N_\textrm{occ}$ matrix. The sign factor merely comes from rearranging fermion creation operators and has no physical effects. Indeed it cancels out when considering the ratio $\langle \textrm{GS} (\Phi_x, 0) \vert \hat{U}_{1;y} \vert \textrm{GS} (\Phi_x, 0) \rangle/\langle \textrm{GS} (0, 0) \vert \hat{U}_{1;y} \vert \textrm{GS} (0, 0) \rangle$ which is precisely the combination appeared in the many-body invariant.

For a fixed $k_x$, we can make a gauge choice so that $u_n (k_x, k_y)$ is smooth as a function of $k_y$ for each $n \in \{ 1, 2, \cdots, N_{\rm occ} \}$~\cite{PhysRev.115.809, PhysRevLett.98.046402}. Then for a large enough $L_y$, we can make the following approximation:
\begin{align}
&\sum_{k_y} \log \bigg( \det \bigg[\Big\langle u_n (k_x + \frac{\Phi_x}{L_x}, k_y) \Big\vert u_m\big(k_x + \frac{\Phi_x}{L_x}, k_y + \frac{2\pi}{L_y} \big) \Big\rangle \bigg] \bigg) \nonumber \\
&= \sum_{k_y} \tr \bigg( \log \bigg[\Big\langle u_n (k_x + \frac{\Phi_x}{L_x}, k_y) \Big\vert u_m\big(k_x + \frac{\Phi_x}{L_x}, k_y + \frac{2\pi}{L_y} \big) \Big\rangle \bigg] \bigg) \nonumber \\
& \approx \sum_{k_y} \tr \bigg( i \Big[\mathcal{A}_{k_y} \Big(k_x + \frac{\Phi_x}{L_x}, k_y \Big)\Big] \frac{2\pi}{L_y} \bigg) = \sum_{n, k_y}  \frac{2\pi i}{L_y} \Big[\mathcal{A}_{k_y} \Big(k_x + \frac{\Phi_x}{L_x}, k_y \Big)\Big]_{(n,n)}  % \approx 2\pi \mathcal{P}_y \Big( k_x + \frac{\Phi_x}{L_x} \Big) \mod 2\pi,
\end{align}
where the Berry connection $\big[ \mathcal{A}_{k_y} (k_x, k_y) \big]$ is an $N_{\rm occ}$-by-$N_{\rm occ}$ matrix whose $(n, m)$th entry is given by $-i \langle u_n(k_x, k_y) \vert \partial_{k_y} u_m(k_x, k_y) \rangle$. Moreover, using the compactness of the Brillouin zone (which is the torus $T^2$), for a sufficiently large $L_x$, one can choose the same smooth gauge for $u_n \Big(k_x + \frac{\Phi_x}{L_x}, k_y \Big) $ and $u_n (k_x, k_y)$, i.e., $u_n$ can be chosen to be smooth on $\big[k_x, k_x + \frac{\Phi_x}{L_x}\big] \times [0, 2\pi]$ for every $k_x$. Hence the many-body invariant reduces to
\begin{align}
\textrm{Im} \bigg[ \log \bigg( \frac{\langle \textrm{GS} (\Phi_x, 0) \vert \hat{U}_{1;y} \vert \textrm{GS} (\Phi_x, 0) \rangle}{\langle \textrm{GS} (0, 0) \vert \hat{U}_{1;y} \vert \textrm{GS} (0, 0) \rangle} \bigg) \bigg] &\approx \sum_{n, k_x, k_y} \frac{2\pi}{L_y} \bigg( \Big[ \mathcal{A}_{k_y} \Big(k_x + \frac{\Phi_x}{L_x}, k_y \Big) \Big]_{(n,n)} - \big[ \mathcal{A}_{k_y} (k_x, k_y) \big]_{(n,n)} \bigg) \nonumber \\ 
&\approx \Phi_x \frac{1}{2\pi} \int_{\rm BZ} d k_x d k_y \tr  \big[\mathcal{F} (k_x, k_y) \big] \nonumber \\
&= \Phi_x C ,
\end{align}
where $\mathcal{F} = \partial_{k_x} \mathcal{A}_{k_y} - \partial_{k_y} \mathcal{A}_{k_x} + i [ \mathcal{A}_{k_x} , \mathcal{A}_{k_y}]$ is the $N_{\rm occ}$-by-$N_{\rm occ}$ matrix-valued Berry curvature. Note that $-\partial_{k_y} \mathcal{A}_{k_x}$ term vanishes under $k_y$ integration due to our gauge choice in $u$ and the trace of $i [ \mathcal{A}_{k_x}, \mathcal{A}_{k_y}]$ vanishes as well. This completes the proof that the many-body invariant reduces to the band Chern number for band insulating states.

% and $\mathcal{P}_y (k_x)$ is the polarization as a function of $k_x$ which is gauge invariant and defined modulo 1. Finally, the many-body order parameter measures the winding number of $\mathcal{P}_y (k_x)$:
% \begin{align}
% &\frac{1}{2\pi} \textrm{Im} \bigg[ \log \bigg( \frac{\langle \textrm{GS} (\Phi_x, 0) \vert \hat{U}_{1;y} \vert \textrm{GS} (\Phi_x, 0) \rangle}{\langle \textrm{GS} (0, 0) \vert \hat{U}_{1;y} \vert \textrm{GS} (0, 0) \rangle} \bigg) \bigg] \approx \sum_{k_x} \bigg[ \mathcal{P}_y \Big( k_x + \frac{\Phi_x}{L_x} \Big) - \mathcal{P}_y ( k_x ) \bigg] \nonumber \\ 
% &\approx \Phi_x \int_0^{2\pi} d k_x \frac{\partial \mathcal{P}_y (k_x)}{\partial k_x} = \Phi_x C ,
% \end{align}
% i.e., the winding number of $y$-polarization as a function of $k_x$ is the Chern number~\cite{PhysRevLett.102.107603}. \BK{reduction to Berry curvature}

\section{$\langle \textrm{GS} \vert \hat{U} \vert \textrm{GS} \rangle$ vs. Path Integral}
Here we briefly review how the ground state overlap can be related to the evaluation of the path integral. Note that this approach has been employed successfully in the previous literature~\cite{PhysRevB.100.245134}. In this paper, we apply this to the Chern and chiral hinge insulators. 

Generically, we can rewrite the overlap into a path integral as follows: 
\begin{align}
\langle \textrm{GS} (\Phi) \vert \hat{U} \vert \textrm{GS} (\Phi) \rangle = \frac{1}{Z_0} \text{Tr} \Big[\hat{U} \cdot e^{-\int^{\beta}_0 d\tau d^d \bm{r} H [A_a^{BC}]}\Big]_{\beta \to \infty},  \label{PathIntegral}
\end{align}
where we have used 
\begin{align}
\vert \textrm{GS}(\Phi) \rangle = \frac{1}{\sqrt{Z_0}} \sum_n \exp\Big(-\frac{1}{2}\int d\tau d^d \bm{r}  H [A_a^{BC}]\Big)|n\rangle,  \nonumber
\end{align}
with $\vert n\rangle$ being the basis of the Hilbert space under the periodic boundary conditions. Here the trace is over $\vert n\rangle$, and $\Phi$ denotes the twist in the twisted boundary condition. For twisted boundary conditions in the spatial directions that we use in the main text, they are implemented by introducing the vector gauge fields in the Hamiltonian. Such gauge field configuration is included by the vector gauge field $A_a^{BC}$ with $a=x,y$ for the two-dimensional cases and $a=x,y,z$ for the three dimensional cases. The gauge fields match the twisted boundary condition, for instance, if $x$-direction is twisted, then $\oint A_x = \Phi_x$. In the numerics, the uniform gauge was used, e.g., $A_x = \Phi_x/L_x$. We also have assumed that the ground state is unique, which is the case for the examples of the current paper. 

Next, in our paper the unitary $\hat{U}$ appearing in Eq.\eqref{PathIntegral} can be written as 
\begin{align}
\hat{U} = \exp \Big[i \int d^d \bm{r} \phi(\bm{r}) \hat{n}(\bm{r}) \Big], 
\end{align}
i.e., the exponent is always proportional to the number operator $\hat{n}(\bm{r})$ weighted with a scalar function $\phi(\bm{r})$. This specific scalar function determines which topology to be measured. Note that, this can be entirely rewritten as 
\begin{align}
\hat{U} = \exp \Big[i \int d^d \bm{r} \int d\tau \phi(\bm{r}) \delta (\tau) \hat{n}(\bm{r}, \tau) \Big]. 
\end{align}
Inserting this expression into the trace, we find that Eq.\eqref{PathIntegral} can be written as 
\begin{align}
\frac{1}{Z_0} \text{Tr} \Big[e^{-\int^{\beta}_0 d\tau d^d \bm{r} (-i \phi(\bm{r}) \delta (\tau) \hat{n}(\bm{r}, \tau) +  H [A_a^{BC}])}\Big]_{\beta \to \infty}.  \label{PathIntegral2}
\end{align}
Writing this way, we can now identify $\phi(\bm{r})\delta(\tau)$ as the time-direction component of the gauge field. Now, we can straightforwardly interpret the above: the trace means that we are performing the evolution of the ground state for all possible configurations of electrons or matter fields of the theory, and hence the result must be the effective theory in terms of $A_\mu$ such that its spatial part is pinned by the boundary condition $A_a^{BC}$ and its temporal component is pinned by the unitary $\phi(\bm{r})\delta(\tau)$. 

Hence, we find that the ground state overlap in Eq.\eqref{PathIntegral} is 
\begin{align}
\langle \textrm{GS} (\Phi) \vert \hat{U} \vert \textrm{GS} (\Phi) \rangle  = |N_\Phi| e^{i\phi_0} e^{i S_{eff}[A_\mu]}, 
\end{align}
where $\phi_0$ is a non-topological constant phase factor. Such factor often appears in similar many-body invariants such as the original Resta's formula.~\cite{PhysRevLett.80.1800} Here $|N_\Phi|$ is the modulus of the overlap and $S_{eff}$ is the effective topological response of the insulators, such as the Chern-Simons term. To remove the effect of $\phi_0$, we take a quotient by $\langle \textrm{GS}(0) \vert \hat{U} \vert \textrm{GS} (0) \rangle = |N_0| e^{i \phi_0}$. This is because we know that the contribution $S_{eff}$ vanishes without the twisted boundary conditions for the cases we consider in this paper. Hence, in total, we find:   
\begin{align}
\frac{\langle \textrm{GS} (\Phi) | \hat{U} \vert \textrm{GS} (\Phi) \rangle}{\langle \textrm{GS}(0) \vert \hat{U} \vert \textrm{GS}(0) \rangle}  = \Big|\frac{N_\Phi}{N_0}\Big| e^{i S_{eff}[A_\mu]}. 
\end{align}
This completes the connection between the ground state overlap and the path integral/topological field theory. With this, we can calculate the phase factors by plugging the gauge field configurations and effective topological field theory, which is straightforward.

\section{Many-body invariant for Infinite Matrix Product State}
The many-body invariants use the ground states on the torus geometry. In interacting 2d systems, such as interacting Chern insulators, it is often useful to use the infinite density matrix renormalization group (iDMRG) to find the ground state in the quasi-1d cylinder geometry. The iDMRG finds the ground state in an infinite matrix product state (iMPS) form. Hence, It is natural to ask if there exits an expression for the many-body invariant, i.e., Eq.(4) of the main text, in terms of iMPS states. In the following, we present such many-body invariant using iMPS states. 

\begin{figure}[t]
\centering\includegraphics[width=0.8\textwidth]{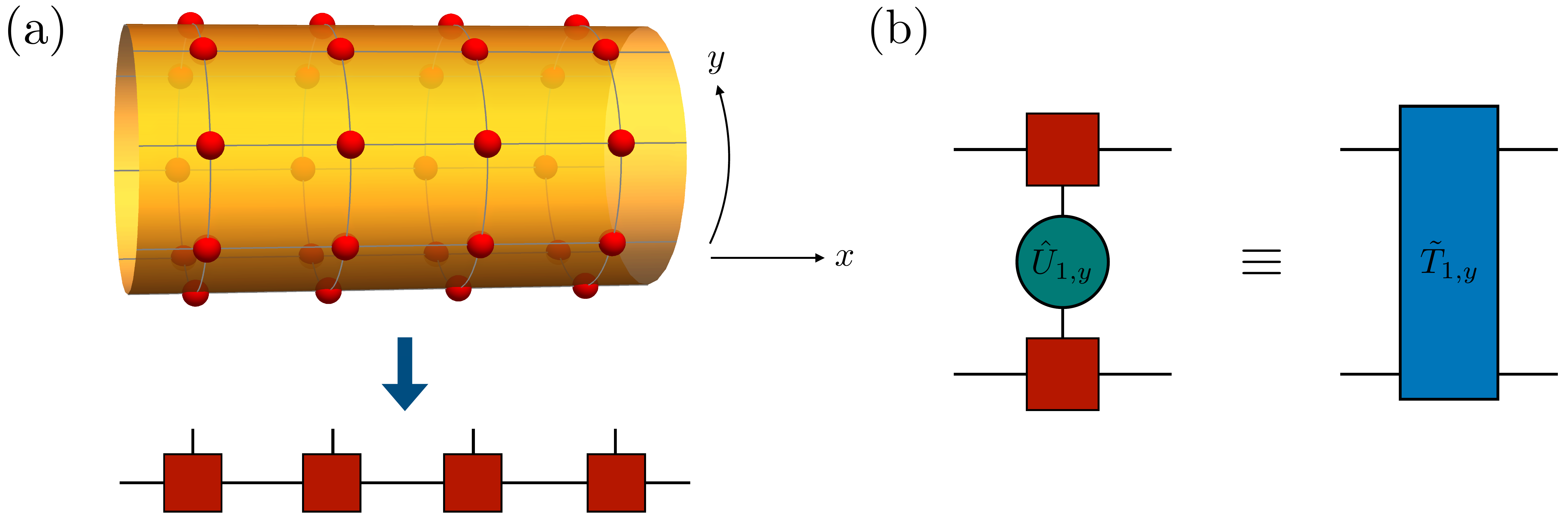}
\caption{(a) We regard a quasi-one dimensional geometry (the cylinderical geometry) used in the iDMRG simulation as a one-dimensional geometry with a `super-site' for each $x$ and consider an ordinary iMPS state for this one-dimensional geometry. (a) By acting $\hat{U}_{1,y}$ operator on the physical bond of a `super-site' MPS, we define the dressed transfer matrix $\tilde{T}_{1,y}$.}
\label{Fig:iMPS}
\end{figure}

Similar to the many-body invariants written in terms of the ground state wavefunctions in the main text, we need two iMPS states obtained from iDMRG simulations on a quasi-1d geometry, one without flux and the other with a constant flux along the infinite direction. We introduce the flux along the infinite direction, the $x$-direction in FIG.~\ref{Fig:iMPS} (a), by coupling the system with a constant gauge field $A_x = \phi_x$. Since we thread the flux along the infinite direction, flux density (flux per unit $x$-length) is infinitesimal, not the full flux, which is divergent. Note that using the constant gauge field $A_x$, the Hamiltonian has the same translational symmetry as the Hamiltonian without the gauge field. Since we are interested in non-fractionalized Chern insulating phases, let us assume that the iMPS ground states also enjoys the translational symmetry along the $x$-direction, i.e., the MPS tensor of the 'super-site' in FIG.~\ref{Fig:iMPS} (a) is independent of $x$ position. Having obtained two iMPS ground states, we measure the polarization along the finite circumference direction, the $y$-direction in FIG.~\ref{Fig:iMPS} (a), using  $\prod_x \hat{U}_{1,y} = \prod_x \exp \big( \frac{2\pi i}{L_y} \sum_{y} y \hat{n}_{x, y} \big)$, where $x$ dependence in $\hat{U}_{1,y}$ is implicit. Note that ``$\hat{U}_{1,y}$" appearing in the main text corresponds to $\prod_x \hat{U}_{1,y}$ in this notation. Although the expectation value of $\prod_x \hat{U}_{1,y}$ may not be well-defined, its asymptotic behavior is governed by the eigenvalue with maximum modulus $\lambda_\textrm{max}^T$ of the dressed transfer matrix $\tilde{T}_{1,y}$, where $\hat{U}_{1,y}$ acts on the physical legs as described in FIG.~\ref{Fig:iMPS} (b). (We assume that the iMPS states are properly normalized, i.e., the usual transfer matrix has $1$ as the eigenvalue with maximum modulus. Also by the assumption of the uniqueness of the ground state, eigenvalue $1$ is non-degenerate.) The complex phase factor of $\lambda_\textrm{max}^T$ corresponds to ($2\pi$ times) $y$-polarization per unit $x$-length, which is well-defined. Finally, the many-body invariant for iMPS states are defined as
\begin{equation}
|\mathcal{Z}_c| e^{i C \phi_x} = \exp \Big( \frac{\lambda_{\rm max}^T (\phi_x)}{\lambda_{\rm max}^T (0)} \Big) ,
\end{equation}
where $|\mathcal{Z}_c|$ is the absolute value of the ``partition function'' and $\lambda_{\rm max}^T (\phi_x)$ is the eigenvalue with maximum modulus of the dressed transfer matrix $\tilde{T}_{1,y}$ for the iMPS ground state of the Hamiltonian with flux $A_x = \phi_x$. 

Note that if we adiabatically change the flux density $\phi_x$ from $0$ to $2\pi$,  the ratio $\lambda_{\rm max}^T (\phi_x)/\lambda_{\rm max}^T (0)$ starts from $1$ and ends at $1$. This implies that the winding number associated with the process, i.e., the Chern number $C$, is quantized. Our claim indicates that it is actually sufficient to look at just a single number $\lambda_{\rm max}^T (\phi_x)/\lambda_{\rm max}^T (0)$ with a nonzero $\phi_x$, not the full winding.

\bibliography{Many-Body}